\documentclass{cernyrep} 
\usepackage{texnames}
\usepackage[T1]{fontenc}
\def\Journal#1#2#3#4{ {#1} {\bf #2} #3 (#4)}

\def\NPA{{\em Nucl. Phys.} A}
\def\PLB{{\em Phys. Lett.} B}

\def\PRL{\em Phys. Rev. Lett.}
\def\PRD{{\em Phys. Rev.} D}
\def\PRC{{\em Phys. Rev.} C}

\def\ZPA{{\em Z. Phys.} A}

\def\ra{\rightarrow}
\def\be{\begin{equation}}
\def\ee{\end{equation}}
\def\bea{\begin{eqnarray}}
\def\eea{\end{eqnarray}}
\def\beas{\begin{eqnarray*}}
\def\eeas{\end{eqnarray*}}

\newcommand{\obb}{0\mbox{$\nu\beta\beta$-decay~}}
\newcommand{\zbb}{2\mbox{$\nu\beta\beta$-decay~}}
\newcommand{\nbb}{neutrinoless double $\beta$-decay~}

\newcommand{\gess}{\mbox{$^{76}$Ge}}

\newcommand{\ndhf}{\mbox{$^{150}$Nd}}
\newcommand{\caav}{\mbox{$^{48}$Ca}}

\newcommand{\xehs}{\mbox{$^{136}$Xe}}

\newcommand{\bec}{\ensuremath{\beta^+/{\rm EC}}}

\newcommand{\ema}{\ensuremath{\langle m_{\nu_e} \rangle}~}
\newcommand{\sint}{\ensuremath{\sin^2 2\theta}}
\newcommand{\delm}{\ensuremath{\Delta m^2}}

\begin{document}
\title{Long term prospects for double beta decay}
 
\author{Kai Zuber}

\institute{Institut f\"ur Kern- und Teilchenphysik, Technische Universit\"at Dresden, 01069 Dresden, Germany}

\maketitle 

\begin{abstract}
In rather general terms the long term perspective of double beta decay is discussed. All important 
experimental parameters are investigated as well as the status of nuclear matrix element issues.
The link with other neutrino physics results and options to disentangle the underlying physics
process are presented.  
\end{abstract}
 
\section{Introduction}
The lepton number violating process of \nbb 
\be
(A,Z) \ra (A,Z+2) + 2 e^- \quad (\obb)
\ee
plays a crucial role in neutrino physics. It can only occur if two conditions are
fulfilled: A neutrino has to be its own antiparticle (called Majorana neutrino) 
AND it has to have a non-vanishing rest mass. For the moment this statement 
ignores any other lepton-number violating
contribution from Beyond Standard Model physics. However, if \obb will ever be
discovered the question will arise how to discriminate the various ideas experimentally.
Nevertheless, Schechter and Valle have shown that the fact of observing \nbb is equivalent
of stating that neutrinos are Majorana particles independent of the dominant mechanism
driving the decay \cite{sch82}. The near term future of experimental activities
over the next five years is covered in \cite{cre09} and thus this article assumes
most of the time that the claimed evidence \cite{kla04} will not be confirmed. 
For the following discussion towards
an ''ultimate'' double beta experiment a very general approach is taken to
work out the important items.\\
One of the major equations governing this field is obviously the radioactive decay law.
It can be written in the case that the half-life  $T_{1/2}$ is much longer than the measuring
time $t$ as
\be 
N_{\beta \beta} = ln 2 \times a \times M \times t \times N_A /T_{1/2}
\ee
with $M$ as the used mass, $a$ the isotopical abundance of the nuclide under interest and
$N_{\beta \beta}$ as the number of double beta decays. How is this linked to experiment?
The experimental signature of
\obb is that the sum energy of the two emitted electrons equals the total nuclear transition energy,
called the Q-value. Thus, in an experiment aiming to see two electrons with a fixed energy 
there are two options, either there are no other physics processes (including the allowed
neutrino accompanied double beta decay \zbb) doing exactly the same
(background free) or you might have some events resulting in the region of interest
(background). These result in different dependences of the expected half-life sensitivity
as
\bea 
\label{eq:hwz}
(T_{1/2})^{-1} \propto a \times M \times \epsilon \times t \quad ({\rm background~ free}) \\
(T_{1/2})^{-1} \propto a \times \epsilon \times \sqrt{\frac{M t}{B \Delta E}} \quad ({\rm background~ limited})
\eea
with $\epsilon$ as the efficiency for detection, $B$ the background index (typically quoted in counts/keV/kg/yr)
and $\Delta E$ as the energy resolution at the peak position. Depending on the (non-)observation of a peak
either a half-life or a lower limit on the half-life can be measured and linked to the neutrino mass via
\be
\label{eq:ema}
(T_{1/2})^{-1} = PS^{0\nu} \times \mid M^{0\nu} \mid^2 \times (\frac{\ema}{m_e})^2
\ee
with $PS^{0\nu}$ as the phase space factor containing Coulomb corrections, $M^{0\nu}$ the nuclear transition
matrix element and the effective Majorana neutrino mass \ema ~defined as
\be
\label{eq:bbquant}
\ema = \mid \sum U_{ei}^2 m_i \mid
\ee
with $U_{ei}$ as the PMNS mixing matrix elements and thus constrained by oscillation results. It should be
noted that due to the condition of requiring neutrinos to be Majorana particles two more CP-violating phases
occur in the mixing matrix, in addition to the one which can be searched for in oscillation experiments.\\
Thus, the ''ultimate'' experiment is easy to define: Infinite measuring time, infinite number of source atoms,
no background, a $\delta$-function for energy resolution, 100 \% efficiency and the largest possible phase space
and nuclear matrix element. The question is how close this can be achieved. As the measuring time is something
which cannot really be effected and most experiments using the ''source=detector'' approach have 100 \% efficiency 
these two factors won't be discussed in the following.

\begin{figure}
\centering\includegraphics[width=7 cm, height= 4cm]{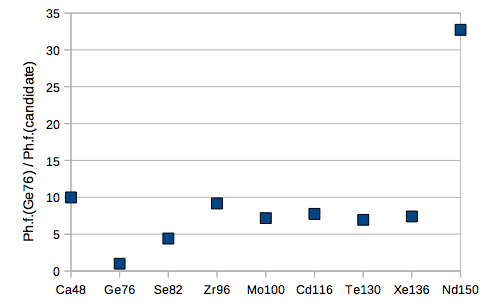}
\centering\includegraphics[width=7 cm, height=4 cm]{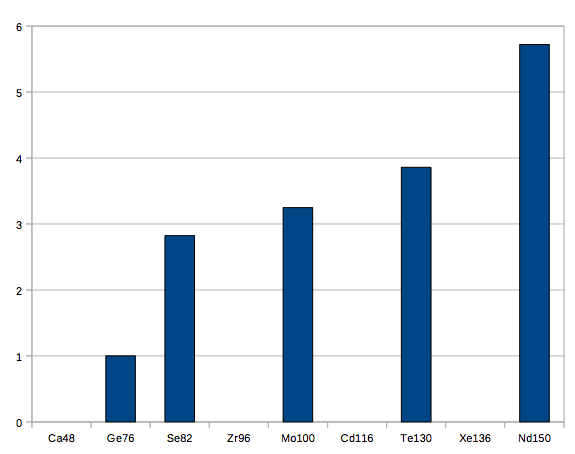}
\caption{Left: Phase space factors including Coulomb corrections for some of the most interesting isotopes, normalise to 
\gess =1. Values taken from \protect \cite{boe92}. Right: Product of phase space and IBM matrix elements \protect \cite{iac09} 
as conversion factor of half-lives into neutrino masses. Again the plot is normalised to \gess =1. }
\label{fig:histo}
\end{figure}

\section{Theoretical arguments}
\begin{figure}
\centering\includegraphics[width=6cm, height=4cm]{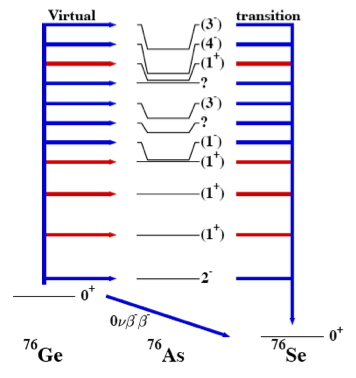}
\centering\includegraphics[width=6cm, height=4cm]{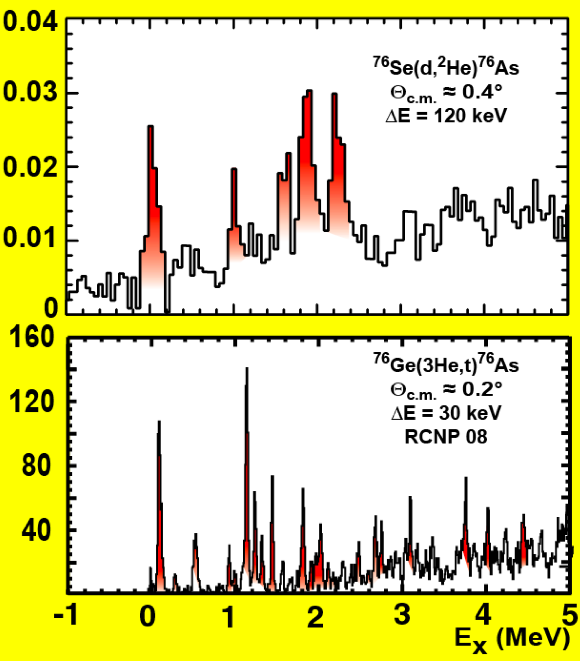}
\caption{Left: Schematic plot of the multipole transitions taking place in double beta decay, as an example \gess~ is used.
In \zbb only the intermediate 1$^+$-states contribute while in the \obb decay all multipoles must be taken into account.
Right: Measurement of 1$^+$-states as measured in charge exchange reactions at RCNP. Multiplying corresponding lines
in both plots and summing up all pairs of line should result in the matrix element reproducing the \zbb half-life (from
\protect \cite{fre09}).}
\label{fig:multipoles}
\end{figure}

The optimal isotope from theory is given by \Eref{eq:ema} , i.e. the one with the largest phase space and Coulomb 
correction combined with the largest nuclear matrix elements. The first quantity is the easiest to determine and histogramed in 
\Fref{fig:histo}.
As can be seen \ndhf~ would come of best. The Coulomb correction supersedes even the phase space dependence
of the Q-value (it scales with Q$^5$ for \obb), the latter would make \caav~ the most suitable nuclide. 
A much more complicated statement involves the nuclear matrix elements.
Their calculation is a severe challenge in nuclear structure physics. In contrast to the \zbb which can be described by 
pure Gamow-Teller transitions through intermediate 1$^+$-states, the neutrinoless mode  can also occur through 
other multipoles. Three different kinds of calculations are used, nuclear shell model calculations (NSM) \cite{cau08}, quasi-particle random
phase approximations (QRPA) \cite{sim09,kor07} and the Interacting Boson Model (IBM) \cite{iac09}. 
To support the calculations an experimental program was launched to provide as much input to the calculations as possible
\cite{zub05}.
Two especially interesting reactions are charge exchange reactions via (d,$^2$He) and ($^3$He,t) and ft-value measurements of
 electron capture and beta decay of the intermediate nucleus. In the first type of reaction the cross section
measurement under zero degrees is directly linked to the Gamow-Teller strength B$_{GT}$ for 1$^+$-states. 
Summing their contribution should reproduce the observed \zbb half-lives (\Fref{fig:multipoles}).
Already quite interesting nuclear structure features have been discovered, which might lead to a deeper understanding of the involved physics.
First measurements under larger angles have been performed, which might reveal also contributions from higher multipoles,
but the relation is not straightforward.\\
In general, various improvements have been achieved in recent years. The problem of short range correlation is much better understood now
and nuclear deformation is about to be included in the matrix element calculations. Past calculations have been done always
for spherical nuclei, first results from studies using deformation show that the general tendency is a reduction in the nuclear
matrix elements and that not to absolute deformation matters, but the shape difference between the mother and daughter
isotope. Deformation is especially important for \ndhf~ as rare earth isotopes are strongly deformed.
Additionally, it has been investigated that the errors in matrix element calculations can be be split into correlated and 
uncorrelated parts \cite{fae09}. This is especially important when comparing different isotopes. For example, typically the quenching is 
included in the axial-vector constant $g_A$ being part of the nuclear matrix element. It is often chosen
to be either $g_A$=1 or 1.25. Obviously changing $g_A$ from one value to another will have the same effect on all isotopes,
thus it would not be correct to compare two nuclides by using the highest matrix element for one and the lowest for the
other.

\section{Background}

Definetly the major experimental effort goes in the reduction of background, i.e. events depositing
the same energy in a detector as \obb . Of course the aim is to get this as low as possible to take advantage of being
''background free'' (first formula in \Eref{eq:hwz}). In this case the half-life sensitivity scales linearly with measuring 
time. Current best background levels achieved are in the order of about 0.1 counts/keV/kg/yr in the \obb peak range, which
is already quite an achievement. However, if the claimed evidence in \gess~ turns out to be wrong then the next 
benchmark is given by the hierarchical neutrino mass models and oscillation parameters, ie. neutrino masses of
about 50 meV. This will require an improvement by 1-2 orders of magnitude in background. The background components can differ
from experiment to experiment, but the typical components
are 

\begin{itemize}
\item The natural radioactive decay chains of U, Th including Rn in the air
\item Radioisotopes produced by spallation processes of cosmic rays
while materials were on the surface
\item Neutron reactions underground produced by fission ($\alpha,n$) or
muon interactions in the rock
\item Muons itself
\item Potentially $^{40}$K as long living isotope
\item The \zbb 
\end{itemize}

\begin{figure}
\centering\includegraphics[width=7cm, height=4cm]{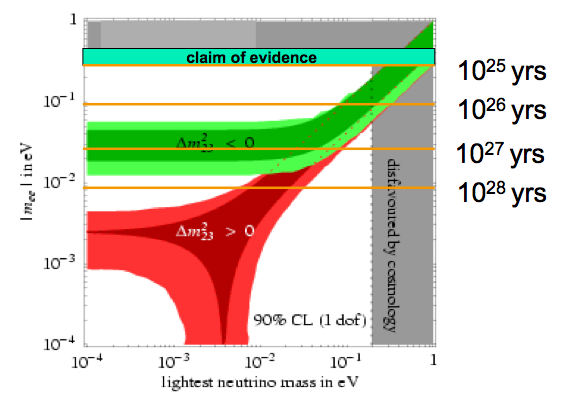}
\centering\includegraphics[width=7cm, height=4cm]{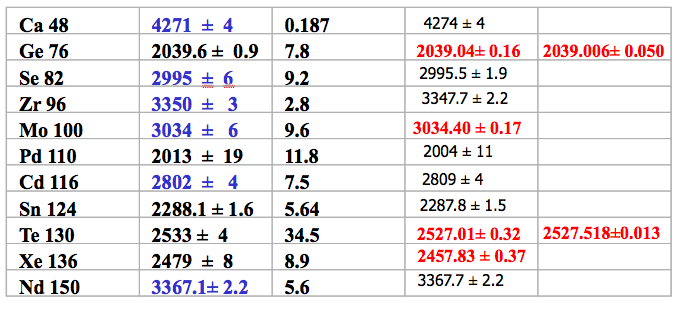}
\caption{Left: Schematic plot of the expected effective Majorana mass as a function of the lightest neutrino mass. Shown is the
range of the evidence and the two hierarchical models. For comparison rough half-life ranges are given. Right: Double beta
emitters with a Q-value of at least 2 MeV. The second column is the Q-value according to \protect \cite{boe92}, the numbers
beyond 2.614 MeV do not suffer from gammas of the natural decay chain. The third column reflects the natural abundance of
the isotope and the last two columns show new Q-values either from the latest atomic mass evaluation \protect \cite{aud03} or as
measured with Penning traps.}
\label{fig:presamples}
\end{figure}

One potential benefit is working with an isotope which has a Q-value above 2.614 MeV, which is the energy of the
highest $\gamma$-line occuring in the natural decay chains (from $^{208}$Tl decay). This implies already 1-2 order
of magnitude lower background to start with and six isotopes are in this preferred range. Material selection is obviously
one of the most important but time consuming activities for every experiment. For the next generation contamination
levels in the region down to $\mu$Bq/kg must be achieved, which requires technologies to measure such small
values. One idea is to electroform copper, a material used in almost all experiments, underground to avoid cosmogenic 
production of $^{60}$Co. Apart from this purely passive methods in minimising impurities in all used materials also
new ideas exist to actively reject background events. This might include pulse shape analysis to discriminate 
single site events (like double beta decay) from multiple site events, tagging the daughter ion created in the
double beta decay in addition to the electrons or aiming for particle identification by using tracking detectors. A good example here is
COBRA, planning to use pixelated CdZnTe semiconductors for this issue. As can be seen in \Fref{fig:presamples}
a clear discrimination between $\alpha$'s, $\beta$'s and muons will be possible.   

\section{Energy resolution}
Another important item is energy resolution. The few expected double beta decay events should be piling
up at the Q-value and not be distributed to widely. This is even true if there is no external background due
to the irreducible component of \zbb. As the measured half-lives of \zbb are typically several orders of magnitude
smaller than the expected one for \obb, thus the fraction of \zbb in the peak range has to be as small as possible
which is directly linked to energy resolution. In this respect, semiconductor detectors, especially Ge-diodes, and 
cryogenic bolometers are leading.

\section{Source mass}
Another crucial item is the used mass, as it determines the source strength. Already with the achieved results 
currently it is obvious that future experiments have to use isotopically enriched material. As can be seen in
\Eref{eq:hwz} even in the background limited case the half-life sensitivity will depend linearly on this quantity.
The process of enriching several hundred kilograms of isotopes is quite expensive and currently dominantly done 
in Russia using ultracentrifuges. Recently, to option of ion cyclotron resonance (ICR) has been explored in some detail.
Its advantage would be its flexibility to enrich various materials. Also the more isotope specific atomic vapour laser
ionisation spectroscopy (AVLIS) could be used. From the financial point of view noble gases are the cheapest to
enrich, thus favouring \xehs~ as a good isotope.

\section{Disentangling the physics process}
If a peak at the Q-value of the transition indeed will be observed, various issues will show up before 
extracting a neutrino mass. The first thing is to probe whether it is really double beta decay and not 
some potential background component. The obvious thing to do is using another isotope. 
\Fref{eq:ema} combined with certain assumptions on the nuclear matrix elements predict a certain half-life 
region. The likelihood that at a different energy in another isotope a background component
will produce the same peak with the correct half-life ratio is very small. If really established that
there is a neutrinoless double beta peak, the question will arise about the underlying physics process.
There are various Beyond the Standard Model processes which would allow for $\Delta L =2$ processes
like right-handed weak currents (V+A interactions), R-parity violating SUSY (the interesting parameter
here is $\lambda_{111}'$), double charged Higgs-bosons and Kaluza-Klein excitations, just to mention
a few. However, the Schechter-Valle theorem \cite{sch82} guarantees that neutrinos are Majorana particles
because at some level of perturbation theory it is always possible to draw a Feynman diagram for
a Majorana mass, but the contribution to \obb is not known. Most of the new particles predicted
in the TeV range hardly change any experimental observable in double beta decay and therefore
constraints of other searches like at the LHC become important to rule out these options. Two mechanisms,
namely KK-excitations and V+A interactions, might be directly testable. Recently it has been shown
that the nuclear matrix elements show a sensitivity to KK-excitations and thus by comparing different
measured isotopes it might be possible to extract information on that \cite{pae07}. Quite a big effect might be caused
by V+A interactions. The single electron energy spectrum and also the opening angle between the 
two emitted electrons is completely different compared to the neutrino mass mechanism,
 it has been shown that an alternative process like
\bec~ has an enhanced sensitivity to V+A interactions \cite{hir94} and also $0^+ \ra 2^+$ should be
dominated by this process. By taking recoil effects into account the latter process might
not be a good candidate for discrimination anymore \cite{tom00}. The process of \bec~ would require
completely new isotopes to be considered, like $^{106}$Cd. An advantage will be the experimental
signature of two 511 keV photons combined with characteristic X-rays, a disadvantage is that the
available transition energy is only $Q-2m_ec^2$ because of the positron creation. The first mentioned
process to discrimate neutrino mass and V+A interactions would require to measure single electron
energy spectra and their opening angle, i.e. a detector with tracking capabilities. Currently only three options of
this kind are discussed, foils of enriched material in a large number of TPCs (SuperNEMO),
filling enriched $^{136}$Xe in a gas TPC (EXO) and high resolution pixelated CdZnTe
semiconductor detectors (COBRA). \\
From all the above mentioned arguments including the uncertainties in the nuclear matrix elements
it becomes apparent that we have to investigate \obb in at least 3-4 isotopes to be able to extract
a reliable neutrino mass value.

\begin{figure}
\centering\includegraphics[width=.3\linewidth]{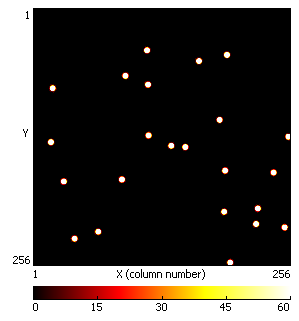}
\centering\includegraphics[width=.3\linewidth]{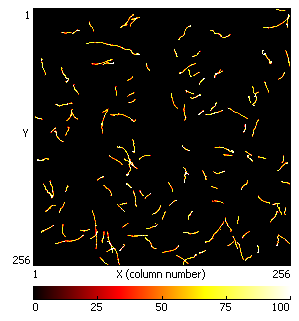}
\centering\includegraphics[width=.3\linewidth]{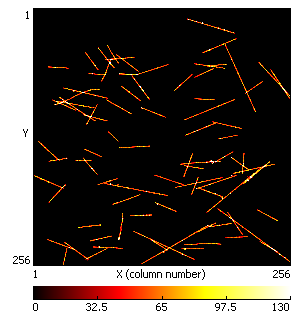}
\caption{Preselected sample of alpha (left) and beta particles (middle) as well
as muons (right) obtained by COBRA with a $55 \mu$m pixel detector. Evident
is the particle identification and the potential tracking option.}
\label{fig:histo}
\end{figure}

\section{Impact of other neutrino measurements}
\begin{figure}
\centering\includegraphics[width=.4\linewidth]{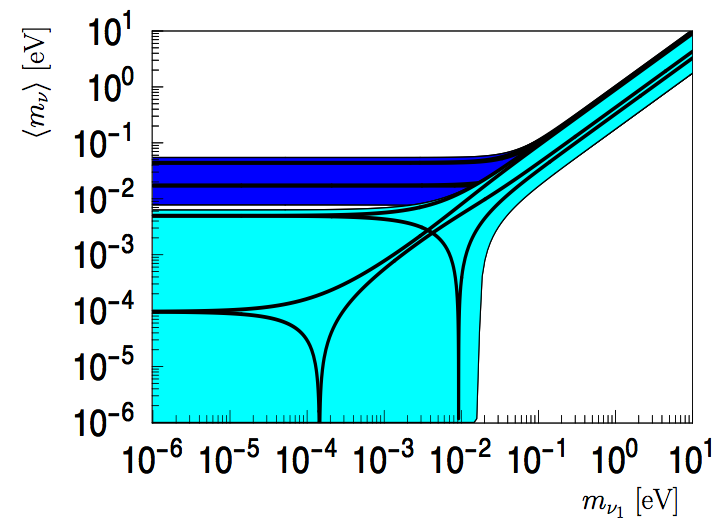}
\centering\includegraphics[width=.4\linewidth]{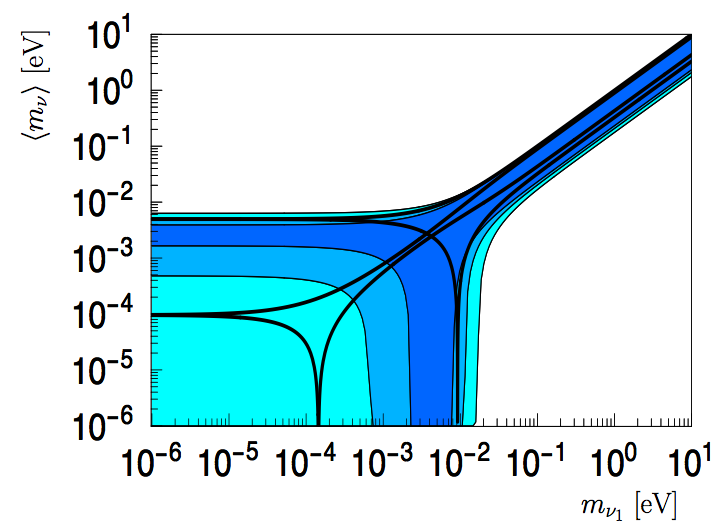}
\caption{Left: Effective Majorana mass as a function of the lightest neutrino mass eigenstate.
Within the given intervalls of oscillation parameters over a wide range a cancellation in the
normal hierarchy can be seen, while the inverted hierarchy has a lower bound due to the
non-maximal mixing of solar neutrinos. Right: The region of cancellations shrinks significantly
if the value of $\sin^2_{13}$ can be restricted to be less than 0.051, 0.025 and 0.0051 respectively. 
Thus, the next generation of reactor experiments and superbeams will help a lot. 
For both plots, the other parameters are fixed to be $\delm_{23} = 1.4 -3.3 \times 10^{-3} eV^2, 
\delm_{12} = 7.2 -9.1 \times 10^{-5} eV^2$ and
\sint$_{12}$ = 0.23-0.38. Nuclear  matrix element uncertainties are not included in
the plot (from \protect \cite{hir06}).}
\label{pic:mahirsch}
\end{figure}

Neutrinoless double beta decay is not the only neutrino physics process
and thus can be linked with other mass measurements onx oscillation results.
Rewriting \Eref{eq:bbquant} in terms of oscillation parameters results in
\be 
\ema = \cos^2_{12} \cos^2_{13} m_1
+ \sin^2_{12} \cos^2_{13} e^{\rm i \alpha} \sqrt{m_1^2 + \delm_{12}} 
+ \sin^2_{13} e^{\rm i \beta} \sqrt{m_1^2 + \delm_{12} + \delm_{23}}
\ee
Thus, the actual predicted value of \ema also depends on an accurate
knowledge of oscillation parameters. For example, assuming CP-invariance
(i.e. the complex phases will be $\pm$1)
and $\sin^2_{13} =0$ can lead to a complete cancellation in the normal hierarchy
scenario if $ \mid {\rm tan} \theta _{12} \mid = m_1/m_2$. An even wider range of
cancellations is possible if the current upper bound on $\sin^2_{13}$
is taken. In the inverted scheme this is not possible as there is a lower
limit due to the non-maximal value of $\theta_{12}$. 
Both facts are shown as an example in \Fref{pic:mahirsch}. 

Furthermore, cosmology provides bounds on the sum of the neutrino masses $\sum m_i$
and beta decay endpoint measurements depend on
\be 
m_\nu^2 = \cos^2_{12} \cos^2_{13} m_1^2
+ \sin^2_{12} \cos^2_{13} m_2^2
+ \sin^2_{13} m_3^2 
\ee
independent of the character of the neutrino. All this various measurements will constrain
the allowed parameter range for neutrino mass and will hopefully lead to a coherent picture by
the end of the day.

\section{Conclusion}
The observation of \obb would be a striking new step beyond the Standard Model and prove that
neutrinos are their own antiparticles. The next
5 years will set the scene for the long term prospects in the field. In this report the options and
critical quantities are discussed in detail not focussing on any specific experiment. As became
apparent there is no unique preferred isotope to use for the potentially most sensitive search.
Actually, due to the various physics processes possible for total lepton number violation and the
uncertainties in nuclear matrix element calculations it is rather mandatory to measure at least
3-4 different isotopes.

\end{document}